\documentclass[12pt]{iopart}
\usepackage{graphicx}
\usepackage{epstopdf}

\begin{document}

\title[Sensitive gravity-gradiometry...]{Sensitive gravity-gradiometry with atom interferometry: progress towards an improved determination of the gravitational constant.}

\author{F. Sorrentino\footnote[7]{also at: Istituto di Cibernetica CNR, via Campi Flegrei 34, 80078 Pozzuoli (NA), Italy}, Y.-H. Lien, G. Rosi, G. M. Tino$^{**}$}

\address{Dipartimento di Fisica e Astronomia \& LENS, Universit\`a di Firenze, INFN Sezione di Firenze, via Sansone 1, I-50019 Sesto Fiorentino (FI), Italy}
\ead{guglielmo.tino@fi.infn.it}
\author{L. Cacciapuoti}
\address{European Space Agency, Research and Scientific Support Department, Keplerlaan 1, 2201 AZ Noordwijk, The Netherlands}
\author{M. Prevedelli}
\address{Dipartimento di Fisica dell'Universit\`a di Bologna, Via Irnerio 46, I-40126, Bologna,  Italy}
\begin{abstract}
We here present a high sensitivity gravity-gradiometer based on atom interferometry. In our apparatus, two clouds of laser-cooled rubidium atoms are launched in fountain configuration and interrogated by a Raman interferometry sequence to probe the gradient of gravity field.
We recently implemented a high-flux atomic source and a newly designed Raman lasers system in the instrument set-up.
We discuss the applications towards a precise determination of the Newtonian gravitational constant $G$.
The long-term stability of the instrument and the signal-to-noise
ratio demonstrated here open interesting perspectives for pushing the measurement precision below
the 100\,ppm level.

\end{abstract}

\maketitle

\section{Introduction}
Matter-wave interferometry has recently led to the
development of new techniques for the measurement of inertial
forces, finding important applications both in fundamental
physics and applied research. The remarkable stability
and accuracy that atom interferometers have reached for
acceleration measurements can play a crucial role for
science and technology. Quantum sensors based on atom interferometry
had a rapid development during the last decade and different
schemes were demonstrated and implemented. Atom
interferometry is used for precise measurements of gravity
acceleration \cite{Kasevich92,Peters99,Mueller08}, Earth's gravity gradient \cite{Snadden98,McGuirk02}, and rotations
\cite{Gustavson97,Gustavson00}. Currently, experiments based on atom interferometry
are in progress to test Einstein's Equivalence Principle \cite{Fray04,Dimopoulos08GR} and to measure the Newtonian gravitational constant G \cite{Bertoldi06,Fixler07,Tino08}, while
experiments on tests of general relativity \cite{Dimopoulos08GR}, for search of quantum gravity effects 
\cite{Tino03,Ferrari06,Camelia09} and for gravitational waves detection
\cite{Tino04,Dimopoulos09} have been proposed.
Accelerometers based on atom interferometry have
been developed for many practical applications including
metrology, geodesy, geophysics, engineering prospecting and
inertial navigation \cite{McGuirk02,Peters01,Leone06,deAngelis09}. Ongoing studies show that the
space environment will allow us to take full advantage of the
potential sensitivity of atom interferometers \cite{Tino07,Turyshev07}.

Our atom interferometer MAGIA (acronym of: Accurate Measurement of G by Atom
Interferometry) was developed for a precise determination of the
Newtonian gravitational constant $G$. The basic idea of the experiment
and some preliminary results are presented in \cite{Fattori03,Bertoldi06,Lamporesi07,Tino08}. 
We recently improved the experimental set-up by implementing a high-flux atomic source based on a 2D-MOT and a newly designed Raman lasers system.

The Newtonian gravitational constant $G$ plays a key role
in the fields of gravitation, cosmology, geophysics, and astrophysics
and is still the least precisely known among the fundamental constants. 
Based on the
weighted mean of eight values obtained in the past few
years \cite{Mohr08}, in 2006 the Committee on Data for Science and
Technology (CODATA) recommended a value with a relative
uncertainty of 100\,ppm. 
  Although $G$ measurements
have improved considerably since 1998\cite{Mohr00}, the available values are still in poor agreement. Indeed, while the
most precise measurements of $G$ have assigned uncertainties lower than 50\,ppm \cite{Gundlach00,Quinn01,Armstrong03,Schlamminger06,Luo09}, the results differ by many standard deviations among each other.
From this point of view, the realization of conceptually
different experiments can help to identify still hidden
systematic effects and therefore improve the
confidence in the final result.
With a few exceptions \cite{Schlamminger06,Schwarz98,Kleinevoss99}, most experiments were performed
using conceptually similar schemes based on suspended
macroscopic masses as probes and torsion balances or pendula
as detectors. 
In our experiment, freely falling atoms act as probes of
the gravitational field and an atom interferometry scheme
is used to measure the effect of nearby well-characterized
source masses. The projected accuracy for MAGIA
 shows that the results of the experiment
will be significant to discriminate between existing
inconsistent values.

The paper is organized as follows:  in section \ref{apparatus} we describe the principle of measurement and the apparatus, with special emphasis on recent upgrades, while in section \ref{results} we give experimental results.


\section{Principle of measurement and experimental apparatus}
\label{apparatus}

The principle of the MAGIA experiment, the scheme of Raman interferometry and its application
to measure $G$ as well as our experimental apparatus have been described in previous papers \cite{Fattori03,Bertoldi06,Cacciapuoti05,Tino08} and references therein. Here we give a brief review, while in sections \ref{2DMOT}, \ref{Raman} and \ref{masses} we describe the recent progress of the experiment.

In our experiment, $^{87}$Rb atoms, trapped and cooled
in a magneto-optical trap (MOT), are launched upwards
in a vertical vacuum tube with a moving optical molasses
scheme, producing an atomic fountain. Near the
apogee of the atomic trajectory, a measurement of their 
vertical acceleration is performed by a Raman interferometry
scheme \cite{Kasevich92}. External source masses are positioned in
two different configurations and the induced phase shift
is measured as a function of masses positions. In order
to suppress common-mode noise and to reduce systematic
effects, a double-differential scheme has been adopted.
The vertical acceleration is simultaneously measured in
two vertically separated positions with two atomic samples,
that are launched in rapid sequence with a juggling
method. From the differential acceleration measurements
as a function of the position of source masses, and from
the knowledge of the mass distribution, the value of $G$ can
be determined.

In a Raman interferometry-based gravimeter, atoms
in an atomic fountain are illuminated by a sequence of 
light pulses which split, redirect, and recombine the atomic
wave packets. The light pulses are realized with two
laser beams, whose frequencies $\omega_1$ and $\omega_2$ are resonant with
the $\Lambda$-type transition of a three-level atom with two lower
states $|a\rangle$ and $|b\rangle$ and an excited state $|e\rangle$. 
The laser beams,
propagating along the vertical $z$-axis in opposite directions, are used to drive two-photon
Raman transitions between $|a\rangle$ and $|b\rangle$. 

Atoms are first prepared in the state  $|a\rangle$. During the interferometer sequence, a $\pi/2$-pulse with duration $\tau=\pi/2\Omega$, $\Omega$ being the two photon Rabi frequency, splits the atom wavefunction into an equal superposition of  $|a\rangle$ and  $|b\rangle$. The interaction with the Raman beams not only modifies the internal state of the atom, but italso results in a momentum exchange by an amount of $hk_{eff} = h(k_1 + k_2)$ ($k_i = \omega_i/c;$  $i = 1,2$) that modifies the atomic trajectories. Successively, a $\pi$-pulse with a duration of $2\tau$ switches back the internal state from  $|a\rangle$ to  $|b\rangle$ and vice versa, re-directing the atomic trajectories. Finally, a $\pi/2$ pulse recombines the atomic packets in the two complementary output ports of the interferometer.
At the end of the interferometer, the probability
of detecting the atoms in the state  $|a\rangle$ is given by
$P_2=(1-\cos\Phi)/2$, where $\Phi$ represents the phase difference
accumulated by the wave packets along the two
interferometer arms. In the presence of a gravity field,
atoms experience a phase shift $\Phi=k_{eff}gT^2$ depending
on the local gravitational acceleration $g$ and on the time interval $T$ between the Raman pulses \cite{Kasevich92}. The
gravity gradiometer consists of two absolute accelerometers
operated in differential mode. Two spatially separated
atomic clouds in free fall along the same vertical axis are
simultaneously interrogated by the same Raman beams to
provide a measurement of the differential acceleration
induced by gravity on the two samples.

\begin{figure}
\begin{center}
\includegraphics[width=0.8\textwidth]{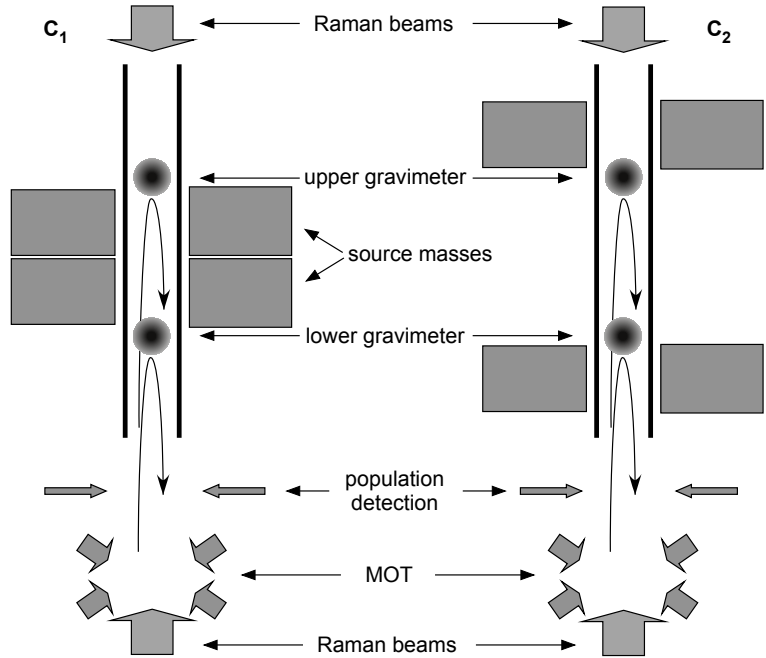}
\caption{\label{magiascheme} Scheme of the MAGIA experiment. $^{87}$Rb atoms, trapped and cooled
in a magneto-optical trap (MOT), are launched upwards
in a vertical vacuum tube with a moving optical molasses
scheme, producing an atomic fountain. Near the
apogees of the atomic trajectories, a measurement of their 
vertical acceleration is performed by a Raman interferometry
scheme. External source masses are positioned in
two different configurations ($C_1$ and $C_2$) and the induced phase shift
is measured as a function of masses positions.}
\end{center}
\end{figure}

Fig. \ref{magiascheme} shows a schematic of the MAGIA experiment. The
gravity gradiometer setup and the configurations of the
source masses ($C_1$ and $C_2$) 
are visible. At the bottom of
the apparatus, a magneto-optical trap (MOT) with beams
oriented in a 1-1-1 configuration collects $^{87}$Rb atoms.
Using the moving molasses technique, the
sample is launched vertically along the symmetry axis of
the vacuum tube and cooled down to a temperature of
 $2.5\,\mu$K. The gravity gradient is probed by two atomic
clouds moving in free flight along the vertical axis of the
apparatus and simultaneously reaching the apogees of their
ballistic trajectories at 60\,cm and 90\,cm above the MOT.
Such a geometry, requiring the preparation and the launch
of two samples with a high number of atoms in a time interval
of about 100\,ms, is achieved by juggling the atoms loaded
in the MOT \cite{Bertoldi06}.
Shortly after launch, the two atomic samples
are velocity selected and prepared in the $(F=1, m_F=0)$
state using a combination of a Raman $\pi$ pulse and resonant blow-away laser pulses. Typically $\sim10^6$ atoms are left after velocity selection. The interferometers take place at the center of the
vertical tube shown in Fig. \ref{magiascheme}. In this region, surrounded by
 two $\mu$-metal shields (76 dB attenuation factor  of the magnetic field
in the axial direction), a uniform magnetic field of 250\,mG along the vertical direction defines the quantization
axis. The field gradient along this axis is lower than
$5\times10^{-5}$\,G/mm. After the Raman interferometry sequence, the population of the ground state is
measured in a chamber placed just above the MOT by
selectively exciting the atoms on the $F=1,2$ hyperfine
levels and sequentially detecting the resulting fluorescence.

Each atom
interferometer in the gravity gradiometer measures the
local acceleration with respect to the common reference
frame identified by the wave fronts of the Raman lasers.
Therefore, even if the phase noise induced by vibrations on
the retroreflecting mirror completely washes out the atom
interference fringes, the signals simultaneously detected on
the upper and lower accelerometers remain coupled and preserve a fixed phase relation. As a consequence, when
the trace of the upper accelerometer is plotted as a function
of the lower one, experimental points distribute along an
ellipse. The differential phase shift is then obtained from
the eccentricity and the rotation angle of the ellipse best fitting
the experimental data \cite{Foster02}.

The source masses \cite{Lamporesi07} are composed of 24 tungsten
alloy (INERMET IT180) cylinders, for a total mass of
about 516\,kg. 
They are positioned on two titanium platforms
and distributed in hexagonal symmetry around the
vertical axis of the tube. Each cylinder is
machined to a diameter of 100\,mm and a height of
150\,mm after a hot isostatic pressing treatment applied
to compress the material and reduce density inhomogeneities.
The two platforms can be precisely translated along the vertical direction by four step motors, with a resolution of 2\,$\mu$m provided by an optical encoder \cite{Lamporesi07}.

The MAGIA apparatus 
has been recently upgraded. The main changes concern the atomic source to load the MOT, the Raman laser system 
 and the shape of source masses.
In the preliminary measurement of $G$ described in \cite{Tino08}, the statistical uncertainty amounted to $1.6\times10^{-3}$ while the systematic uncertainty amounted to $4.6\times10^{-4}$.  
The major contributions to the systematic uncertainty budget came from the knowledge of the position of the source masses with respect to the atomic trajectories and of the atomic initial velocity. 

In the present work, we have addressed both the systematic uncertainty and the long term stability to show that they are compatible with a measurement of $G$ at the level of 100 ppm. 
We discuss improvements on source mass positioning in section \ref{masses} while an estimate of the uncertainty on atomic velocity is discussed in section \ref{velocity}.

\subsection{\label{2DMOT}2D-MOT}

In the experiment reported in \cite{Tino08}, the magneto-optical trap was loaded from the background Rb vapour obtained from a dispenser.
A major disadvantage of such approach was the obvious trade-off between MOT loading rate and background pressure in the vacuum system. Indeed, a high Rb vapour density is important for a fast MOT loading, but it also degrades the vacuum inducing higher atom losses along the all interferometer sequence and more background fluorescence at detection. 

\begin{figure}
\begin{center}
\includegraphics[width=0.7\textwidth]{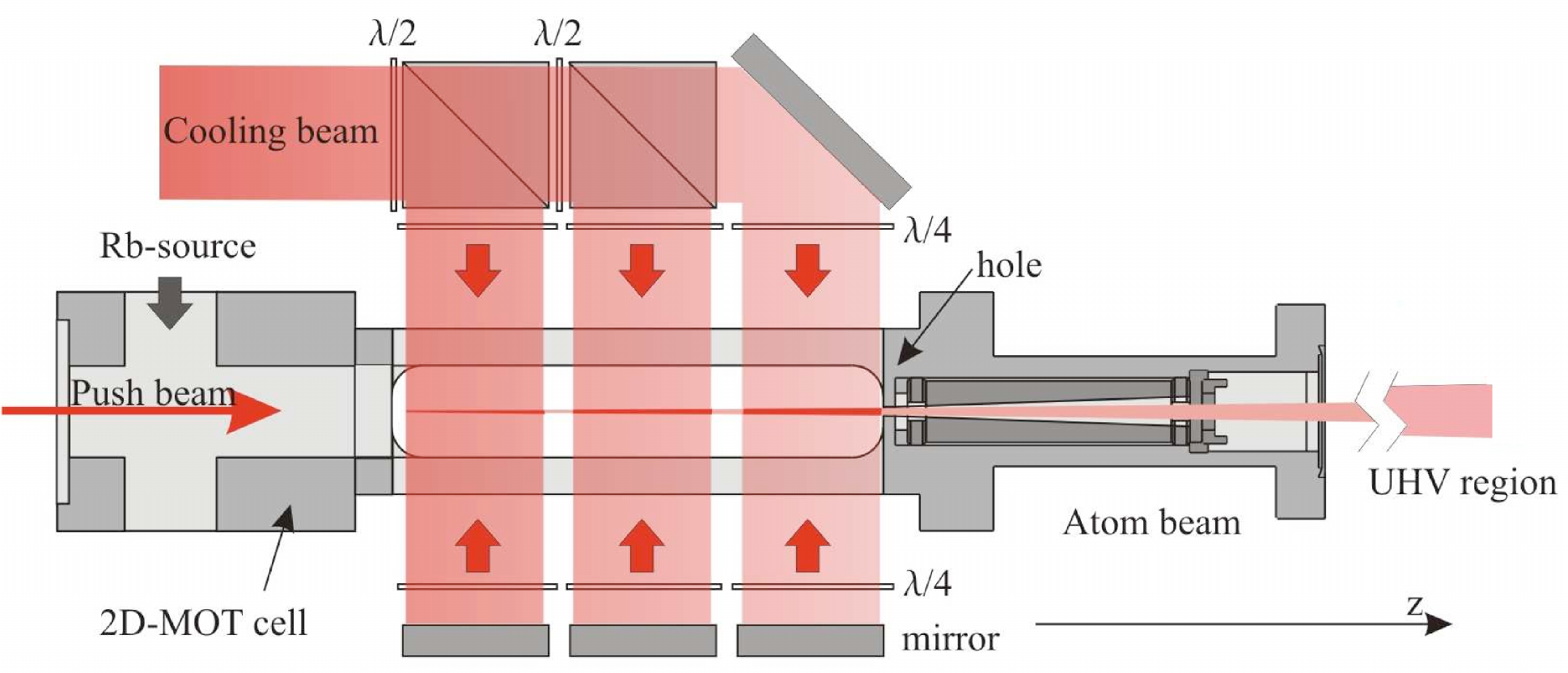}
\caption{\label{2DMOTsystem} 
Scheme of the 2D-MOT system.
}
\end{center}
\end{figure}

To achieve fast loading rates while preserving a very low background pressure in the MAGIA vacuum system, a high flux atomic source based on a two-dimensional magneto-optical trap (2D-MOT)  \cite{Dieckmann98} has been implemented. Atoms evaporate from a temperature controlled rubidium reservoir and, through a 15 mm diameter
tube, they enter a vapor cell with dimensions of $25\times25\times90$\,mm (see fig. \ref{2DMOTsystem}). The cell
is machined from a single piece of titanium and 4 rectangular windows ($15\times80\times3$\,mm) are glued to its
sides for optical access. Two sets of coils are attached to the cell to provide the desired radial
magnetic gradients of about 20\,Gauss/cm. Two orthogonal beam pairs of cooling
laser Ð with repumper overlapped Ð enter the vapor cell through the rectangular
windows and radially cool the atoms. For sake of compactness of the optical setup each beam is split into three circularly polarized parts with 24.5\,mm beam-diameter. A low intensity laser beam, slightly red detuned from the $F=2\to F'=3$ cooling transition and propagating along the axial direction, pushes the atoms increasing the flux in the direction of the 3D-MOT chamber. 
As a result, an atomic beam is coupled out through  a hole (1.5\,mm diameter) at the back wall of the cell (2\,mm thick).
Before entering the UHV chamber,  whose center
is at about 0.5\,m distance from the 2D-MOT, the generated atomic beam passes through a tube of purified graphite with a
conical hole for differential pumping. 

The laser system used to operate the 2D-MOT 
is based on a home-made Master Oscillator - Power Amplifier (MOPA) with output power of about 500\,mW. The master is an extended cavity diode laser using an interference filter for wavelength selection \cite{Baillard06}. Two double-pass AOMs allow independent frequency and power tuning of the cooling and pushing beams. Optimal atomic flux from the 2D-MOT source is found when the optical frequencies of such beams are red tuned from the  $F=2\to F'=3$ transition by 8 and 13\,MHz, respectively.

Under typical operating conditions, measured values for the atomic beam flux, mean axial velocity, velocity spread and atomic beam divergence are $10^{10}$\,atoms/s, 15\,m/s, 7\,m/s,  and 23\,mrad, respectively.

\begin{figure}
\begin{center}
\includegraphics[width=0.6\textwidth]{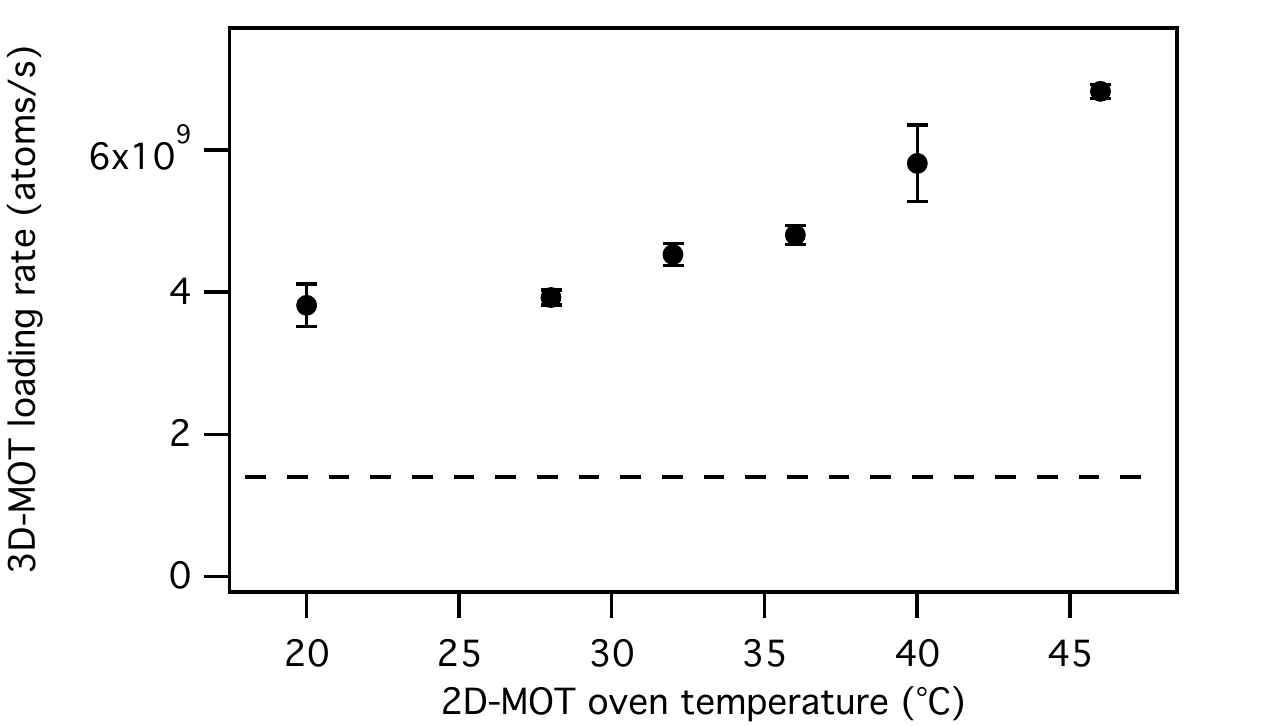}
\caption{\label{2DMOTresults} 
Loading rate of 3D-MOT from 2D-MOT versus temperature of Rb oven; the dashed line indicates the typical loading rate when the 3D-MOT is loaded from the background Rb vapor created by a dispenser.}
\end{center}
\end{figure}

As compared to the standard operating conditions of the dispenser previously employed to load the 3D-MOT, when using the 2D-MOT the background Rb density is reduced by more than two orders of magnitude  in the UHV chamber. Fig. \ref{2DMOTresults} 
shows typical values of the MOT loading rate versus the temperature of the Rb oven in the 2D-MOT system. As compared to the standard operating conditions of the dispenser previously employed to load the 3D-MOT, the MOT loading rate can be increased by up to a factor 5. Since the signal to noise ratio in the interferometer scales as the square root number of atoms, this change is expected to improve the sensitivity of our gravity gradiometer by about a factor 2.2.

\subsection{Raman laser system}
\label{Raman}

\begin{figure}
\begin{center}
\includegraphics[width=0.8\textwidth]{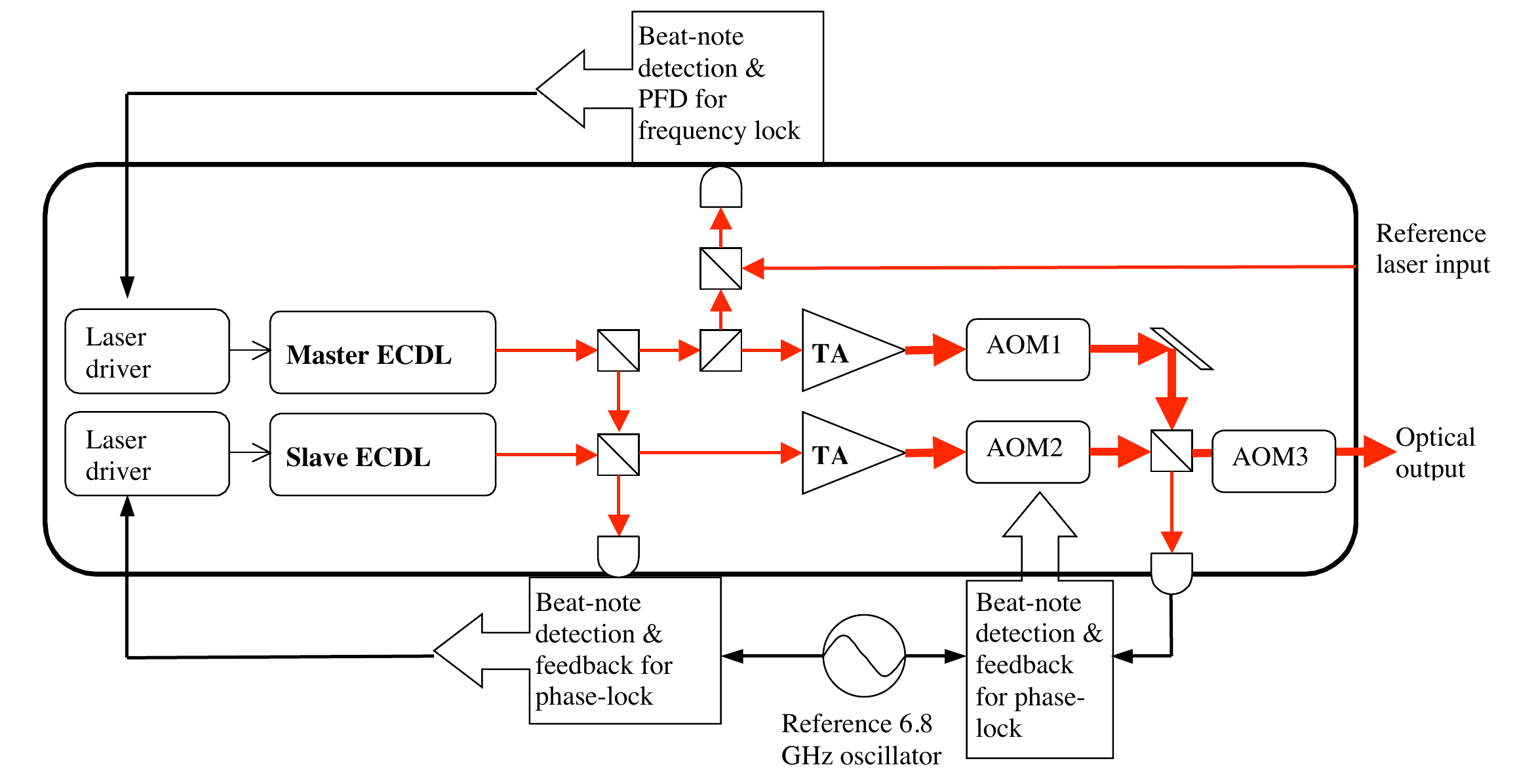}
\caption{\label{ramanscheme} Scheme of Raman laser system.}
\end{center}
\end{figure}

The Raman beams are generated by two home-made MOPA systems with output power of about 700\,mW each (see fig. \ref{ramanscheme}). Two interference filter stabilized extended cavity diode lasers are phase-locked with an offset frequency of about 6.8 GHz generated by a microwave synthesizer. In addition, one of the two lasers (master laser)  is frequency locked with an offset of about 2 GHz to the $F=2\to F'=3$ transition, by detecting the beat note with a frequency stabilized laser (reference laser). Each laser beam injects an indipendent Tapered Amplifier (TA). Such system has several advantages as compared to the apparatus previously employed in the experiment.
Indeed, interference stabilized extended cavity diode lasers have lower intrinsic frequency noise than Littrow grating stabilized lasers, sensibly improving the locking stability and the mean time between unlock events.
In addition, using two independent tapered amplifiers instead of a single one allows independent control on the intensity of the two Raman beams. Fluctuations in the Raman beams intensity might cause drifts in the measured gravity gradient through the light shift effect. Such scheme also provides higher optical power for the Raman beams. The higher Rabi frequency has two effects: it results in a higher efficiency of Raman transitions; moreover, the Raman lasers interact with a larger class of atomic velocities. 
In our setup, the overall optical power of Raman beams after the optical fiber is about 200\,mW.

 Our Raman laser system features a double-stage optical phase-locked loop (OPLL). The primary OPLL detects the beat note between the two ECDL beams before injection of the TA, in such a way to minimize the signal propagation delay and to maximize the loop bandwidth. We mix the beat note with the 6.8 GHz reference frequency and we compare the downconverted signal with a reference frequency from a Direct Digital Synthesizer (DDS)  in a fast digital phase-frequency detector (Motorola MC100EP140). The DDS frequency is swept around 40 MHz with a linear frequency ramp to compensate for the change in Doppler effect during the interferometric sequence. The resulting error signal is properly filtered and used to drive two actuators on one of the ECDL (slave Raman laser), namely, the PZT holding the output coupler and the injection current of the laser diode. The loop bandwidth on the injection current is about 4\,MHz. The output beams from the TAs are passed through two AOMs for independent intensity control, and are finally recombined in a polarizing beam splitter. A third, single-pass AOM is used for pulse shaping just before coupling the Raman beams into an optical fiber.

\begin{figure}
\begin{center}
\includegraphics[width=0.6\textwidth]{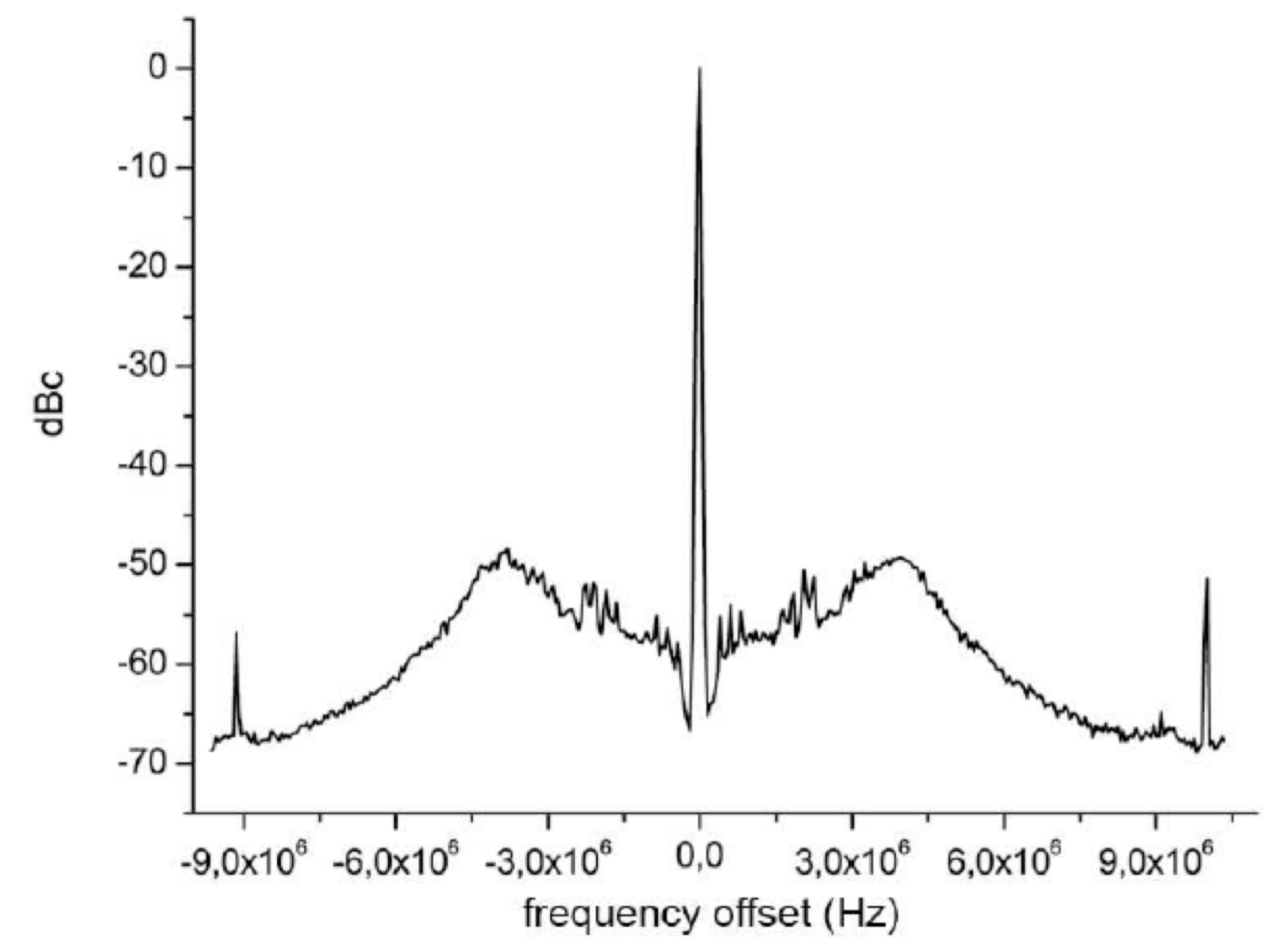}
\caption{\label{beatnote} Beat note between Raman lasers; resolution bandwidth: 30 kHz.}
\end{center}
\end{figure}

\begin{figure}
\begin{center}
\includegraphics[width=0.8\textwidth]{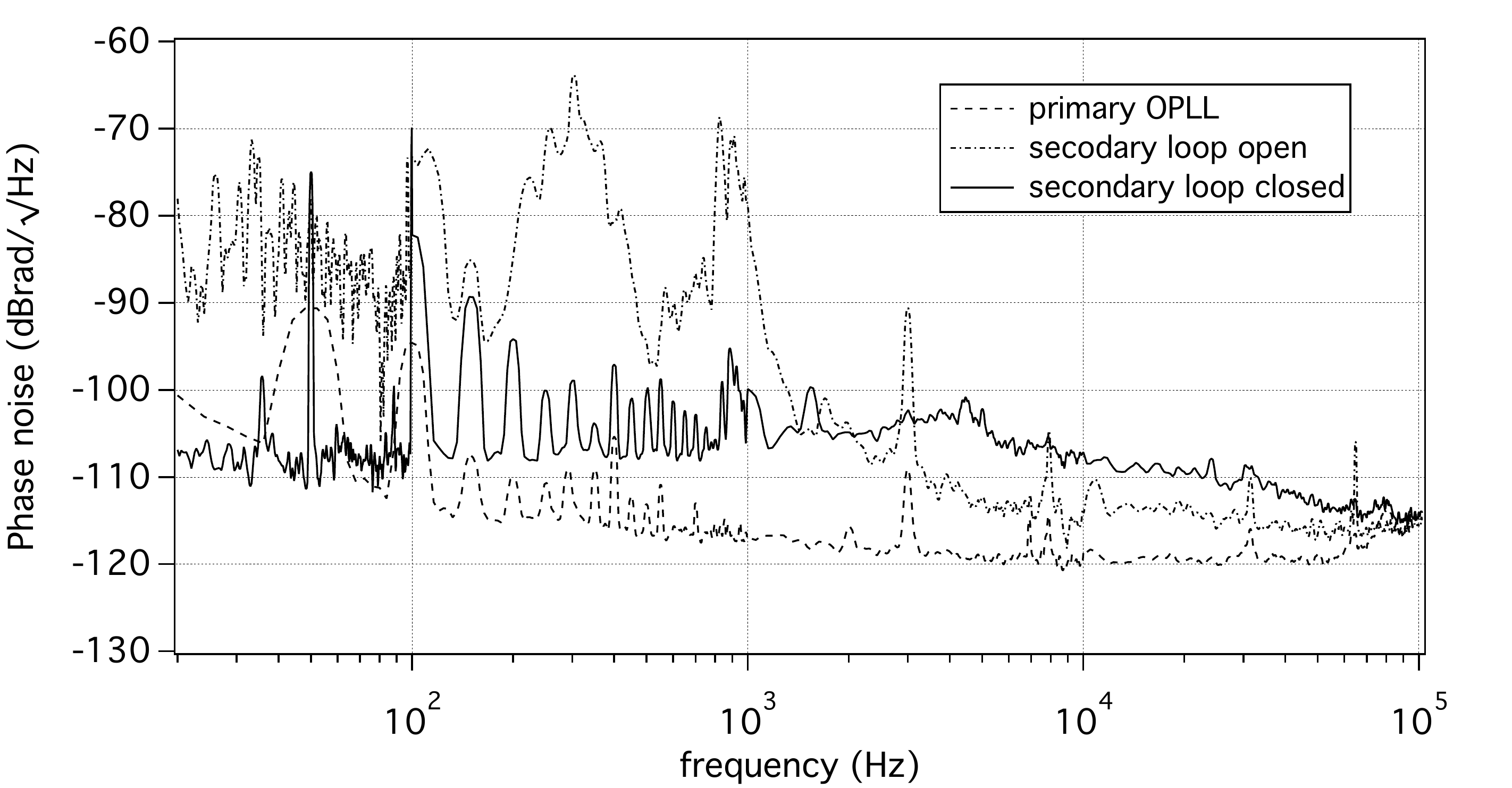}
\caption{\label{phasenoise} Phase noise spectrum of the beat signal between the two Raman laser beams measured in different conditions: after the primary loop (dashed curve); after the optical fiber with the secondary loop open (dash-dotted curve); after the optical fiber with the secondary loop closed (solid curve).}
\end{center}
\end{figure}

We also apply an auxiliary, low bandwidth loop in order to compensate for the phase noise introduced through the differential path of the two Raman laser beams before they are recombined in the optical fiber; to such purpose we detect the beat note between the Raman laser beams at the polarizing beam splitter before the third AOM; we mix the beat note with the 6.8 GHz reference frequency and we use another fast digital phase-frequency detector to compare the downconverted signal with the same $\sim40$\,MHz reference frequency employed in the primary loop. The resulting error signal is properly filtered and used to control a voltage controlled crystal oscillator (VCXO) driving the frequency of the AOM after the TA of the slave Raman beam. The resulting loop bandwidth is about 100\,kHz. Fig. \ref{beatnote} shows the RF spectrum of the beat note between Raman laser, while fig. \ref{phasenoise} shows the phase noise spectral density measured in different conditions: after the primary loop, after the optical fiber with the secondary loop open, after the optical fiber with the secondary loop closed.

\subsection{Source masses}
\label{masses}

In the preliminary measurement of $G$ described in \cite{Tino08}, one of the major contributions to the systematic uncertainty budget came from the knowledge of the position of the source masses with respect to the atomic trajectories; the corresponding contribution to the error on $G$ amounted to $3.6\times10^{-4}$. While an estimate of the uncertainty on atomic trajectories is discussed in section \ref{velocity}, we recently improved the precision of mass positioning in two ways. First, we applied a polishing and rectification machining to the 24 cylinders. After such process, the shape of each cylinder is regular within $5\,\mu$m rms. Then, we tested the use of a laser tracker to measure the position of each cylinder. Since the shape of the cylinders is regular, the position is known once the center and tilt angle are measured. We place the target corner cube of the laser tracker on the upper face of the cylinder under analysis, and we measure its coordinates in several positions on the surface. We fit the measured data to a plane, thus obtaining the elevation and tilt angle of the cylinder. In the final configuration, the horizontal coordinates of the cylinder are measured by placing the corner cube on a small conical mark machined at the exact center of the plane surface. Our tests show that we can measure the relative position of our cylinders with a rms error below 5\,$\mu$m. 
As a result, the contribution of source mass positioning to the systematic relative uncertainty on $G$ can be reduced to the level of $2\times10^{-5}$.


\section{Experimental results} 
\label{results}

In order to characterize the apparatus, we tested the sensitivity 
of the gravity gradiometer and its stability on the time scale of a few days. We also investigated the atomic velocity to improve the systematic uncertainty in the measurement of $G$.

\subsection{Gradiometer sensitivity} 
\label{sensitivity}

\begin{figure}
\begin{center}
\includegraphics[width=0.6\textwidth]{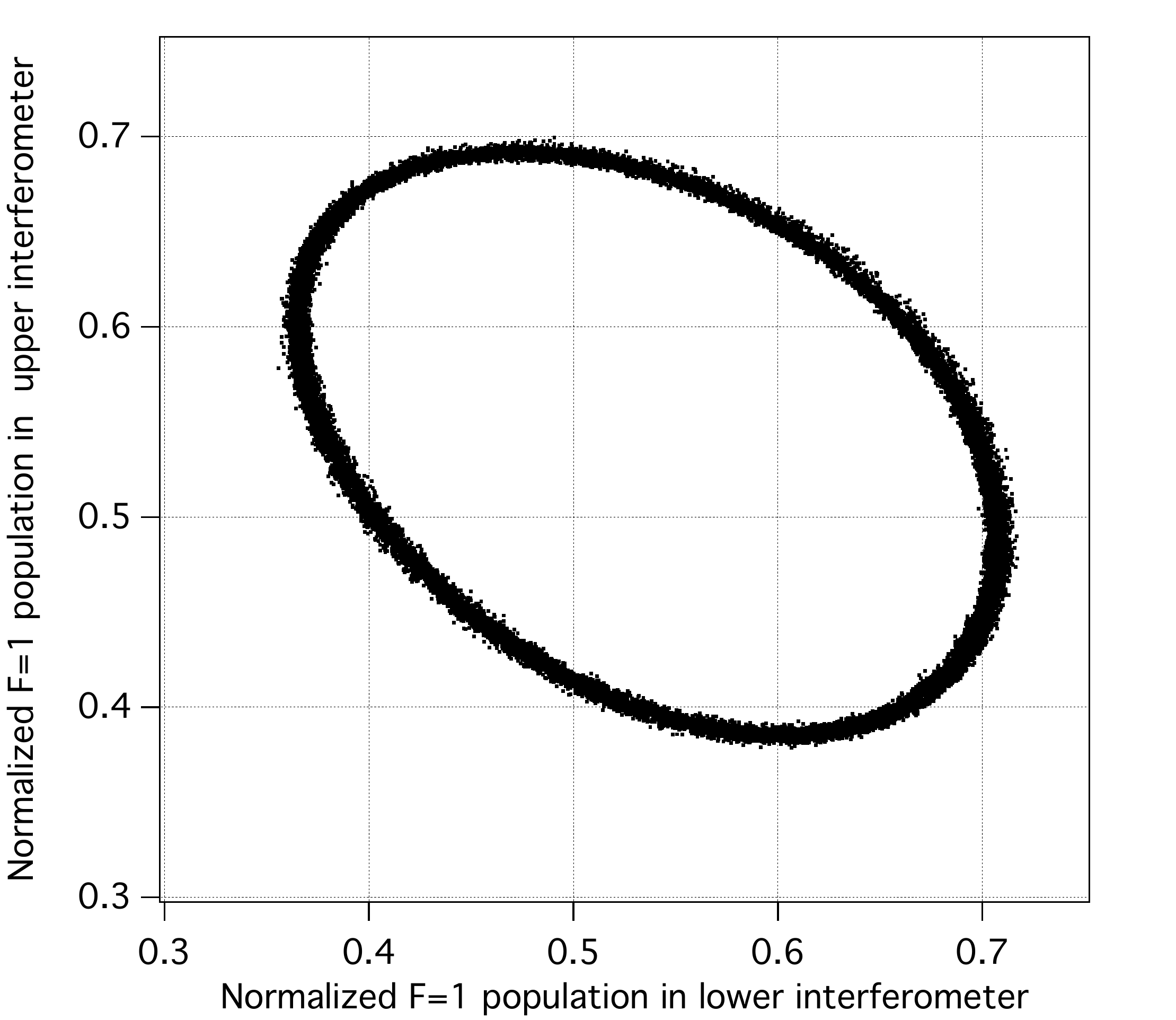}
\caption{\label{25000points} Lissajous plot of 25200 data points.}
\end{center}
\end{figure}

As a first test of the sensitivity of our apparatus, we observed the  statistical fluctuations of the gradiometer measurements over about 17 hours, keeping the masses in a fixed position. Fig. \ref{25000points}  shows a Lissajous figure obtained by plotting the normalized 
population of the $F=1$ ground state detected at the output port of the upper interferometer as a function of the same measurement performed at the output port of the lower interferometer \cite{Bertoldi06}. Each point 
 can be used to extract the two phases modulus $2 \pi$ of the two interferometers
after a single experimental cycle. The repetition time of the experiment is about 2.5\,s, and the plot contains about 25200 points.

The data have been divided in a series of 36 consecutive data points. Each
group of 36 points was fitted with an ellipse and the value of the angle has been
extracted with its estimated error. The optimal number of points required to fit an ellipse has been estimated by varying the number of points $n$ in
the ellipse, evaluating the whole fit, and computing the Allan deviation for the
series. As a figure of merit we consider
$\eta(n)=\sigma(1)\sqrt{n}$, where $\sigma(1)$ is the Allan deviation at 1 ellipse, i.e. $\eta$ is the equivalent Allan deviation for a single point.
 Since the fit is heavily nonlinear $\eta  (n)$ drops sharply at first, then it reaches a plateau and finally starts to increase due to long term
drifts. By choosing $n$ as the smallest value that reaches the plateau both $S/N$
and temporal resolution are optimized.
With $n = 36$, we have 700 complete ellipses from the data
shown in fig. \ref{25000points}. 

\begin{figure}
\begin{center}
\includegraphics[width=0.65\textwidth]{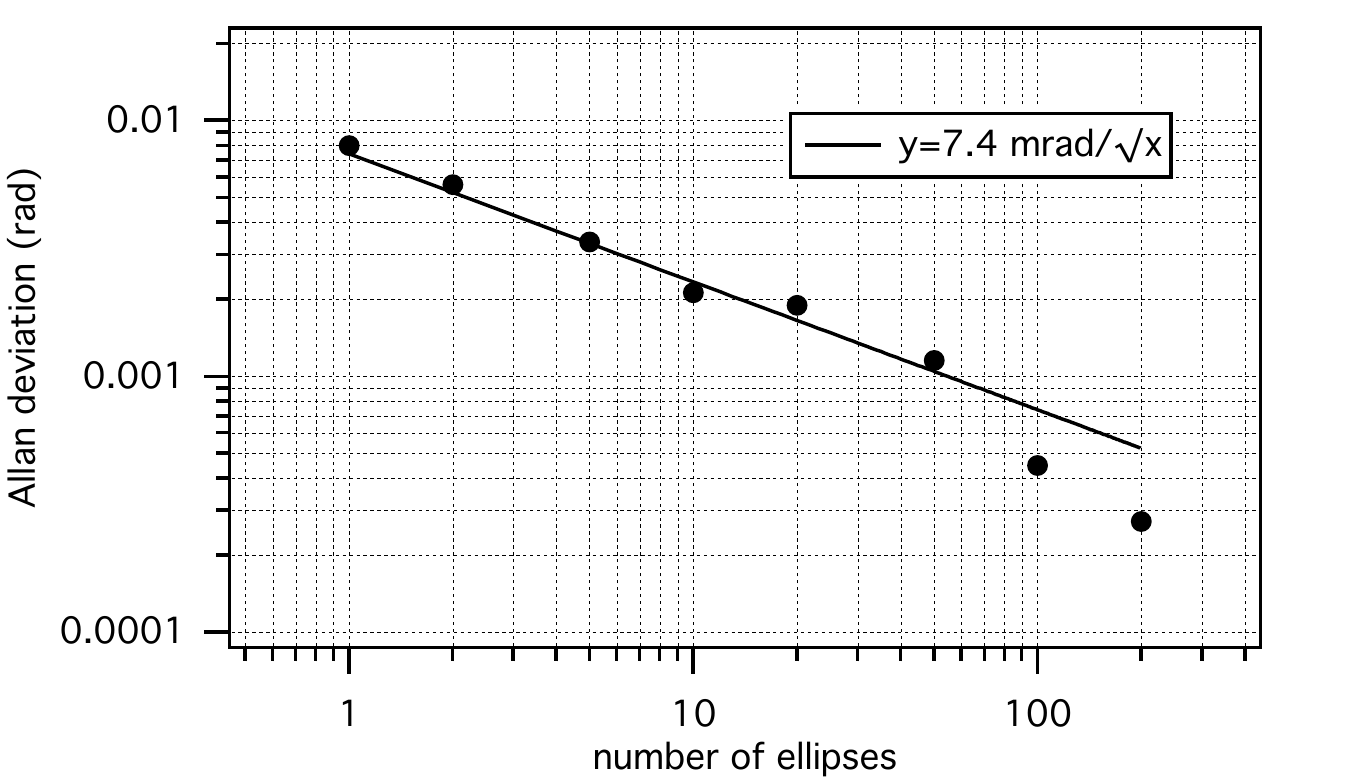}
\caption{\label{allan} Allan deviation of the 
ellipse angles calculated from the data shown in fig. \ref{25000points}.}
\end{center}
\end{figure}

We have evaluated the Allan variance of the
differential phase shift and verified that it scales as the
inverse of the square root of  the integration time, showing the typical behavior
expected for white noise (see fig. \ref{allan}). The instrument has a sensitivity
of 70\,mrad at 1\,s, corresponding
to a sensitivity to differential accelerations of  $1.7\times10^{-8}$\,$g$ at 1\,s. The resulting sensitivity is about a factor two better than in \cite{Tino08}, thus reducing by a factor 4 the integration time needed to reach a specific precision target in the $G$ measurement. The regime of  100\,ppm uncertainty can now be reached in about 40 days of continuous measurement.

\begin{figure}
\begin{center}
\includegraphics[width=0.7\textwidth]{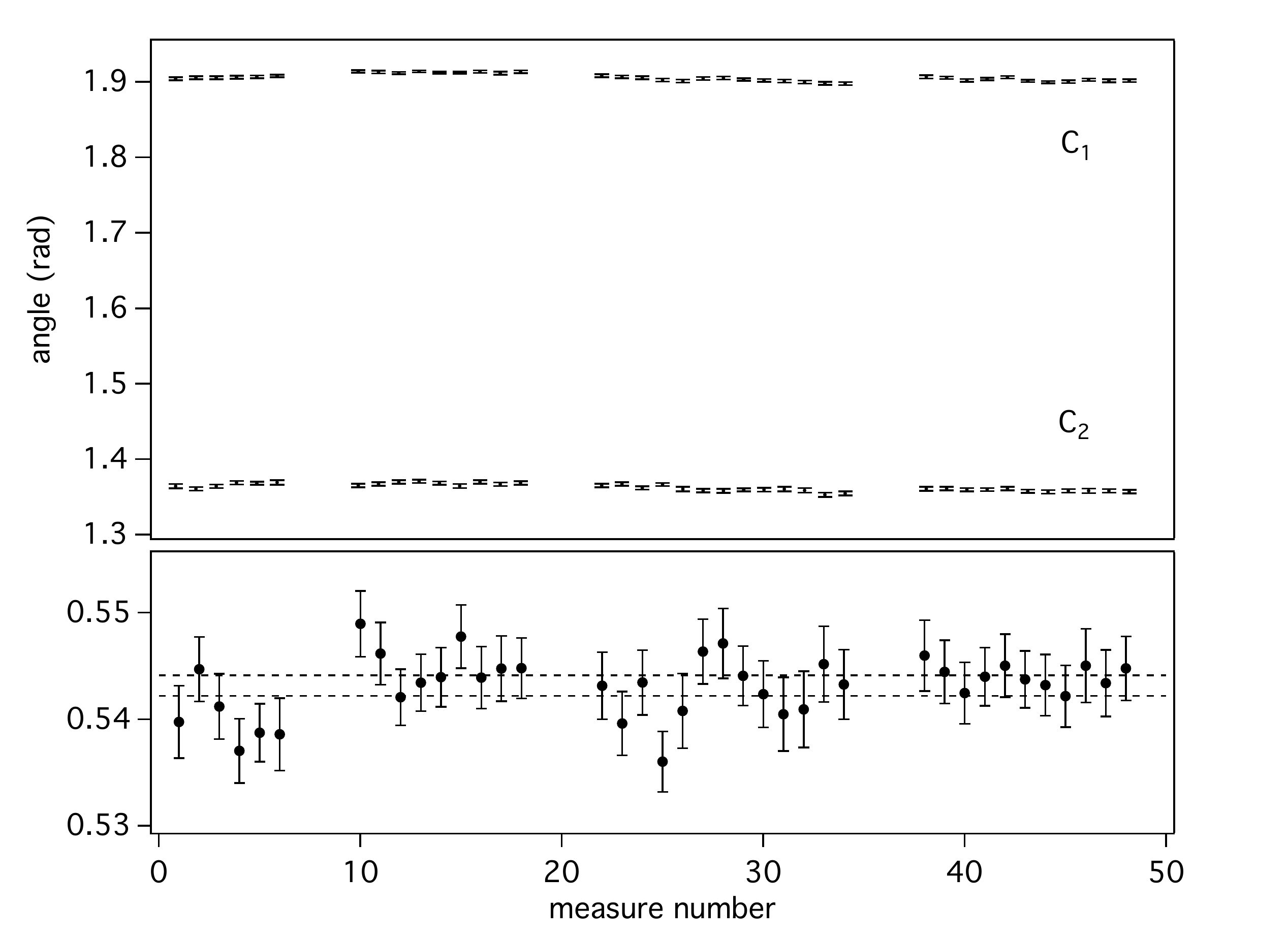}
\caption{\label{movemasses} Modulation
of the differential phase shift measured by the atomic gravity
gradiometer when the distribution of the source masses is
alternated between configuration $C_1$ (upper points) and $C_2$
(lower points). Each point in the upper graph is the weighted average of 40 consecutive phase measurements obtained by fitting
a 24-point scan of the atom interference fringes to an ellipse. The lower graph shows the resulting values of  the angle of rotation $\Phi(i)$; dashed horizontal lines are marking the $\pm2\sigma$ confidence interval.}
\end{center}
\end{figure}

We also modulated the position of the source masses as shown in fig. \ref{magiascheme}. Fig. \ref{movemasses} shows a measurement of the differential interferometric phase on a period of 60 hours over four consecutive days.
We moved the masses from the close ($C_1$) to the far ($C_2$) configuration and viceversa every 40 minutes, corresponding to 960 measurement cycles. We split each set $C_n(i)$ of 960 points ($n=1,2$) in 40 series of 24 consecutive points; we fitted each group of 40 ellipses and we evaluated the average $\Phi_n(i)$ and the standard error on the average $\delta\Phi_n(i)$ of the differential phase.
In the sequence $C_1(i-1), C_2(i), C_1(i+1)$ the linear drift in time was removed by comparing $\Phi_1(i)$ with the weighted average of $\Phi_2(i-1)$ and $\Phi_2(i+1)$. In this way we obtained 39 couples of data $\{\Phi_1(i),\Phi_2(i)\}$. From each couple a value
for the angle of rotation $\Phi(i)$ can be obtained. The final result is  $\Phi=0.54315 \pm 0.00048$\,rad and the $\chi^2$ is 33. This is equivalent to a statistical uncertainty of  $8.8\times10^{-4}$ on the measurement of $G$.


\subsection{Measurement of atomic velocity}
\label{velocity}

In the preliminary measurement of $G$ described in \cite{Tino08}, the contribution to the error on $G$ coming  from the knowledge of the atomic initial velocity amounted to $2.3\times10^{-4}$.
In order to improve the error budget accounting for systematic effects on the $G$ measurement, we have refined the measurement of  the vertical velocity of the atomic sample after velocity selection. 

The measurement technique is based on Raman velocimetry.
We apply a Raman $\pi$ pulse 350\,ms after the velocity selection pulse.
We change the frequency ramp on the Raman beams (see section \ref{Raman}) to successively reach the resonant condition on the $\pi$ pulse corresponding
to the two possible configurations of the wave vector (upwards and downwards oriented).
So the frequency difference between the two resonant peaks gives the
mean velocity of the atomic cloud after velocity selection.
We change the final frequency of the ramp to sweep through the two resonances.

\begin{figure}
\begin{center}
\includegraphics[width=0.8 \textwidth]{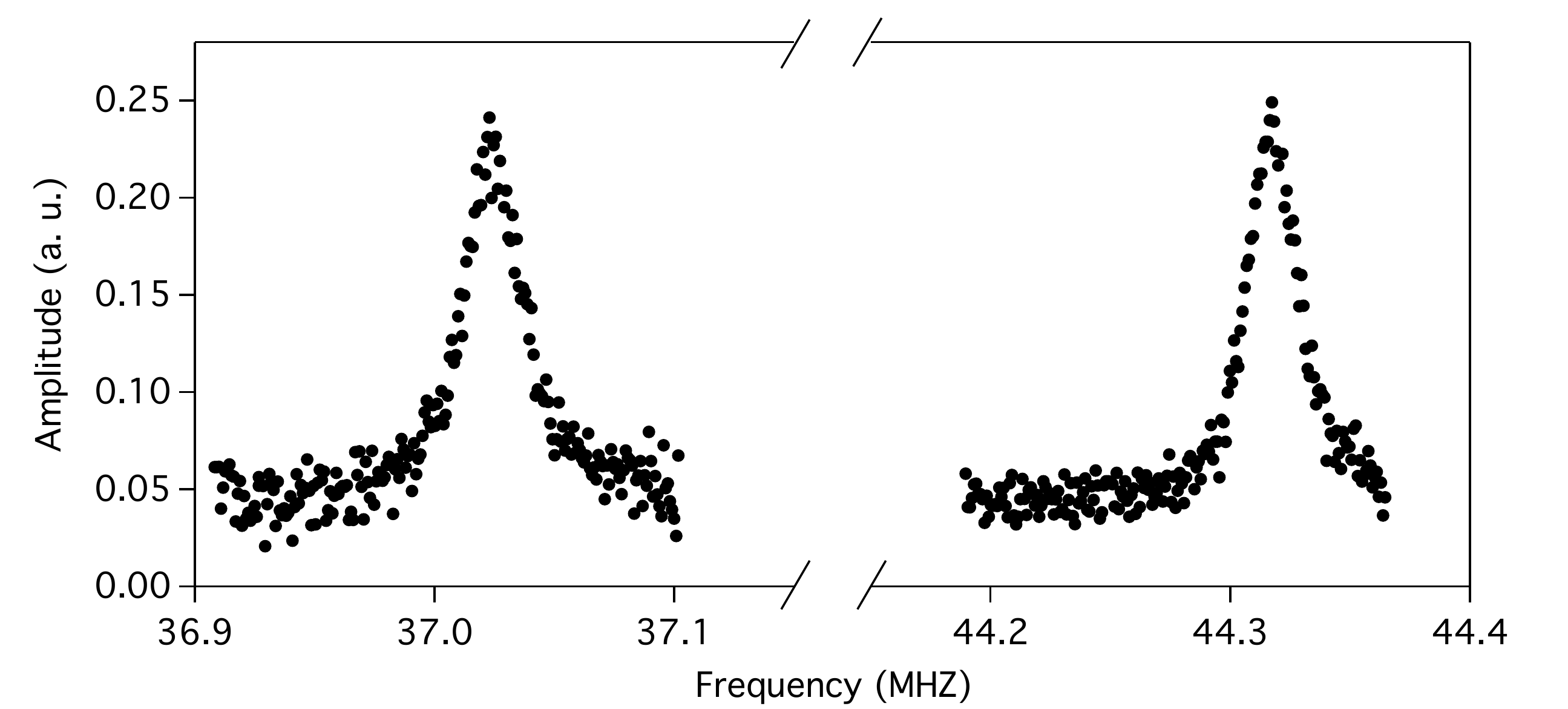}
\caption{\label{velocimetry} Raman velocimetry; 
the graph shows the two resonance peaks of the Raman velocimetry; the second $\pi$ pulse is applied 350\,ms after the velocity selection pulse.}
\end{center}
\end{figure}

Experimental data showing the resonant peaks are presented in fig. \ref{velocimetry}. By fitting the data with Lorentzian shapes we measure a frequency difference between the two peaks of $\Delta\nu=7.30218\pm0.00029\,\hbox{MHz.}$
 Such frequency difference is related to twice the Doppler shift: $2\pi\Delta\nu=2k_{eff}v.$ 
Thus the atomic vertical velocity is measured with a precision $\delta v/v=3.9\times10^{-5}$.
As a result, the contribution of the error on vertical atomic velocity to the systematic uncertainty on $G$ can be reduced below $10^{-5}$.

The degree of control of atomic velocity and source mass position is compatible with a measurement of $G$ with a target accuracy of 100\,ppm. In such conditions, the systematic uncertainty may be limited by the knowledge of the atomic positions relative to the source masses, which is much less critical than source masses positioning in the MAGIA experiment. Because of the high density of tungsten, the
gravitational field produced by the source masses is able to
compensate for the Earth's gravity gradient. Therefore, operating the
interferometers close to stationary points of the gravitational potential strongly reduces the
uncertainty on $G$ due to the knowledge of the atomic
positions \cite{Tino08}. A knowledge of the atomic positions with millimeter precision (both vertically and radially)
will be required to reduce the systematic uncertainty on $G$ below $10^{-4}$.

\section{Conclusions}
\label{conclusions}

We presented 
a sensitive gravity gradiometer based on Raman atom interferometry. 
Recent upgrades to the MAGIA apparatus have allowed to reach a sensitivity to differential gravity accelerations of $1.7\times10^{-8}$\,$g$/s. 

We also discussed the system performance for a measurement of the Newtonian gravitational constant.
Our apparatus can run continuously for several days,
showing a reproducibility of the gravity gradient measurement compatible with 
the stated sensitivity on such time scale. Our measurement of differential gravity gradient over four days is equivalent to a statistical uncertainty of  $8.8\times10^{-4}$ on the measurement of $G$.

We have refined the error contribution deriving from the main sources of systematic uncertainty in the experiment. 
In particular, the effect
 of both the knowledge of source masses positions and vertical atomic velocities on the $G$ error budget can be reduced 
well below 100\,ppm. 

In the next future, a measurement of $G$ with atom interferometry at the level of 100\,ppm seems possible.
With the demonstrated sensitivity, an integration time of
about 40 days will be required for reaching such uncertainty level. Further improvements in the setup, such as the use of more sophisticated detection schemes \cite{Biedermann09} or the implementation of high momentum beam splitters \cite{Mueller09}, may enable an even higher sensitivity.

\section*{References}

\end{document}